\documentclass[preprint,groupedaddress,amsmath,amssymb,prb]{revtex4}
\usepackage{longtable}
\usepackage{lscape}
\usepackage{fancyheadings}
\usepackage{amsmath}
\usepackage{amsfonts}
\usepackage{delarray}
\usepackage{graphicx}
\usepackage{dcolumn}
\usepackage{bm}

\newcommand {\apgt} {\ {\raise-.5ex\hbox{$\buildrel>\over\sim$}}\ }
\newcommand {\aplt} {\ {\raise-.5ex\hbox{$\buildrel<\over\sim$}}\ }

\begin{document}

\title{{First Principles Phase Diagram Calculations
for the Octahedral-Interstitial System ZrO$_{X}$,
$0 \leq X \leq 1/2$}}

\author{Benjamin Paul Burton}
\email{benjamin.burton@nist.gov}
\affiliation{Materials Measyrement Laboratory,
Metallurgy Division, National Institute of Standards and Technology (NIST),
Gaithersburg, MD 20899, USA}
\altaffiliation{Phone: 301-975-6053, FAX: 301-975-5334.}

\author{Axel van de Walle}
\affiliation{Engineering and Applied Science Division,
California Institute of Technology,
1200 E. California Blvd.
MC 309-81
Pasadena, CA 91125;
avdw@alum.mit.edu}

\author{Axel van de Walle}
\affiliation{Engineering and Applied Science Division,
California Institute of Technology,
1200 E. California Blvd.
MC 309-81
Pasadena, CA 91125;
avdw@alum.mit.edu}

\author{Harold T. Stokes}
\affiliation{ Department of Physics and Astronomy, 
Brigham Young University,Provo, Utah 84602, USA;
stokesh@byu.edu}

\date{\today}

\begin{abstract}
First principles based phase diagram calculations were performed for the
octahedral-interstitial solid solution system
$\alpha ZrO_{X}$~
($\alpha Zr[~~]_{1-X}O_{X}$; [~~]=Vacancy; $0 \leq X \leq 1/2$).
The cluster expansion method was used to do a ground state analysis,
and to calculate the phase diagram. The predicted diagram has four
ordered ground-states in the range $0 \leq X \leq 1/2$, but one of these,
at X=5/12, is predicted to disproportionate at T$\approx 20K$, well below
the experimentally investigated range T$ \approx 420K$. Thus, at
T$ \apgt 420K$, the first-principles based calculation predicts
three ordered phases rather than the four that have been reported by
experimentalists.
\end{abstract}

\maketitle

~~~
~~~\\

\section{Introduction}

Zircalloy is used as nuclear fuel-rod cladding in light water
reactors, but it is metastable with respect to oxidation
by the UO$_2$ fuel.\cite{Cronenberg1978,Hoffman1984a,Hoffman1984b,Hoffman1985}

Oxidation of zircalloy transforms it from the high-temperature
(high-T), oxygen-poor, bcc solution ($\beta$Zr$_{X}$) into the low-T,
oxygen-rich, hcp-based solution ($\alpha$ZrO$_{X}$).
At temperatures between about 1173K and 573K various ordered phases
have been reported.
\cite{Holmberg1961,Yamaguchi1968,Fehlmann1969,Yamaguchi1970,Hirabayashi1972,Hirabayashi1974,Hashimoto1974,Arai1976,Sugizaki1985}

Octahedral interstitial ordering of oxygen (O),
and vacancies ([~]) in $\alpha ZrO_{X}$
($\alpha Zr[~~]_{1-X}O_{X}$, $0 \leq X \leq 1/2$)
increases microhardness \cite{Dubertret1966}
and brittleness \cite{Cronenberg1978}, and therefore,
promotes stress corrosion cracking.
Order-disorder transitions were studied via heat capacity
measurements: Arai and Hirabayashi \cite{Arai1976} studied
alloys with O/Zr ratios of 0.16 and 0.24 at 473K-973K;
Tsuji and Amaya \cite{Tsuji1995} made similar measurements on
alloys with O/Zr ratios of 0.0, 0.10, 0.13, and 0.24, at
325K-905K.

Arai and Hirabayashi \cite{Arai1976} achieved a high
degree of long-range ordering in samples that were cooled
from 623K to 523K, during a period of about one month,
which indicates a high mobility of oxygen
in $\alpha Zr[~~]_{1-X}O_{X}$, even at such modest
temperatures; hence a system that is highly reactive
even at such moderate temperatures.

A recent computational study \cite{Ruban2010} reported
ground-state structures and order-disorder transition temperatures 
for Zr$_{6}$O and Zr$_{3}$O, but did not present a calculated phase 
diagram, or report if the calculated order-disorder transitions 
in Zr$_{6}$O and Zr$_{3}$O are first-order, as experiment indicates, 
or continuous.

The results presented below are mostly consistent with
experimental studies with respect to the comparison between
computationally predicted ground-state (GS) structures and reported
(experimental) low-temperature ($T \aplt 500K$)
ordered phases. with the exception that in the range
$0.25 \aplt X \aplt 0.5$ the calculations predict only two
ordered phases at $T \> 150K$, rather than the three called
$\alpha_2^{\prime \prime}, ~ \alpha_3 ^{\prime \prime}$ and $\alpha_4 ^{\prime \prime}$~ in Arai and
Hirabayashi (1976).\cite{Arai1976}

Experimental values for the maximum solubility of O in Zr, X$_{max}$,
range from: $X_{max} \approx $ 29 at. \% \cite{Domagala1954,Yamaguchi1968};
to $X_{max} \approx $ 35 at. \% \cite{Abriata1986}; and
$X_{max} \approx $40 at. \% \cite{Hirabayashi1974,Arai1976,Sugizaki1985}.
The first-principles results presented here support
a higher value; i.e. X$_{max} \geq 1/2$.  This may reflect an
insufficiently negative calculated value for the formation
energy of monoclinic ZrO$_2$.

\section{Methodology}

\subsection{Total Energy Calculations}

\begin{figure}[!htbp]
\begin{center}
\includegraphics[width=10.6cm,angle=0]{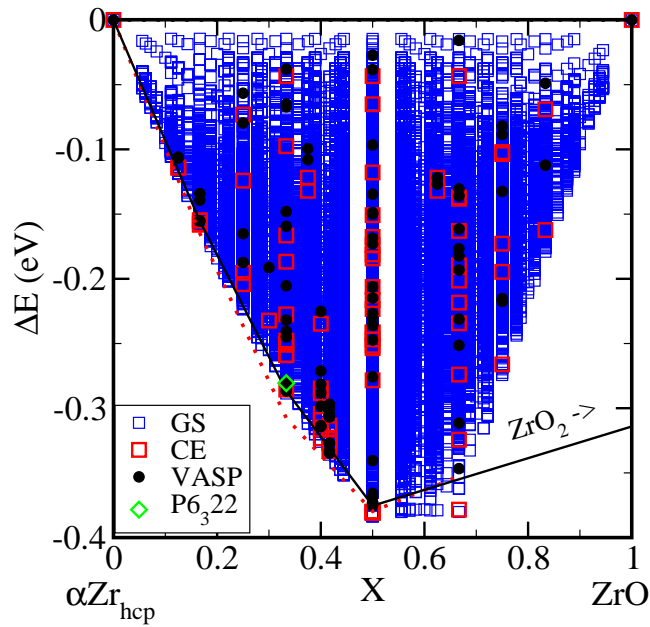}
\end{center}
\vspace{-0.5in}
\caption{Comparison of VASP (solid circles)
and CE (larger open squares, red online)
formation energies,$\Delta E_{f}$, and a
ground-state analysis on structures with 18
or fewer octahedral-interstitial sites
(smaller open squares, blue online).
Extension of the convex hull towards the formation
energy of monoclinic zirconia, ZrO$_2$,
indicates that the four ordered GS
at X=1/6, 1/3, 5/12 and 1/2 are also predicted
to be GS of the Zr-O binary.
}
\label{fg:GS}
\end{figure}

Formation energies, $\Delta E_{f}$~ (Fig. \ref{fg:GS}) were calculated
for fully relaxed hcp $\alpha$Zr, hcp $\alpha$ZrO
(hcp $\alpha$Zr with all octahedral
interstices occupied by O), and 96
$\alpha Zr[~~]_{1-n}O_{n}$~ supercells of intermediate composition.
All calculations were performed with the density functional
theory (DFT) based Vienna $ab~initio$~
simulation program (VASP, version 445 \cite{Disclaimer,Kresse1993}) using
projector-augmented plane-wave pseudopotentials, and the generalized gradient
approximation for exchange and correlation energies.
Electronic degrees of freedom were optimized with a conjugate gradient
algorithm, and both cell constant and ionic positions were fully relaxed.
Pseudopotential valence electron configurations were:
Zr$_{sv}$:~4s4p5s4d; O$_s$:~3s$^{2}$3p$^{4}$.

Total energy calculations were converged with respect
to k-point meshes by increasing the density of
k-points for each structure until convergence.
A 500 eV energy cutoff was used, in the "high precision"
option which guarantees that {\it absolute}~ energies are
converged to within a few meV/site
(a few tenths of a kJ/site of exchangeable species;
O, [~~]).  Residual forces were typically 0.02 eV or less.

Calculated formation energies, $\Delta E_{f}$,
relative to a mechanical
mixture of $\alpha$Zr + $\alpha$ZrO, for the 96
$\alpha Zr[~~]_{1-n}O_{n}$~
supercells are plotted as solid circles in Fig. \ref{fg:GS}.
Values of $\Delta E_{f}$~ are,

\begin{equation}
\Delta E_{f} = (E_{Str} - E_{\alpha Zr} - E_{\alpha ZrO} )/(2)
\end{equation}

\noindent
where: $E_{Str}$~ is the total energy of the
$\alpha Zr[~~]_{1-n}O_{n}$~ supercell;
$E_{\alpha Zr}$ is the energy/atom of $\alpha Zr$;
$E_{\alpha ZrO}$ is the energy/atom of $\alpha ZrO$.

\subsection{The Cluster Expansion Hamiltonian}

The cluster expansion, CE \cite{Sanchez1984}, is a compact representation
of the configurational total energy. In the
$\alpha Zr[~~]_{1-X}O_{X}$ system,
the solid solution configuration is described by pseudospin occupation
variables $\sigma _{i}$, which take values $\sigma_i=-1$~ when site-$i$ is
occupied by [~~] and $\sigma_i=+1$~ when site-$i$ is occupied by O.

The CE parameterizes the configurational energy, per exchangeable cation, as
a polynomial in pseudospin occupation variables:

\begin{eqnarray}
E(\mathbf{\sigma })=\sum_{\ell }m_{\ell }J_{\ell }\left\langle
\prod_{i\in \ell ^{\prime }}\sigma _{i}\right\rangle
\end{eqnarray}

\noindent
Cluster $\ell$~ is defined as a set of lattice sites. The sum is taken over all
clusters $\ell$~ that are not symmetrically equivalent in the
high-T structure space group, and  the average is taken
over all clusters $\ell ^{\prime }$ that are
symmetrically equivalent to $\ell$.
Coefficients $J_{\ell}$~ are called effective cluster interactions, ECI, and
the \emph{multiplicity} of a cluster, $m_{\ell}$, is the
number of symmetrically equivalent clusters, divided by the number of cation sites.
The ECI are obtained by fitting a set of VASP FP calculated
structure energies, $\{ E_{Str} \}$.
The resulting CE can be improved as necessary by increasing the
number of clusters $\ell$~ and/or the number of $E_{Str}$~ used in the fit.

\begin{figure}
\begin{center}
\vspace{-0.5in}
\includegraphics[width=8.0cm,angle=0]{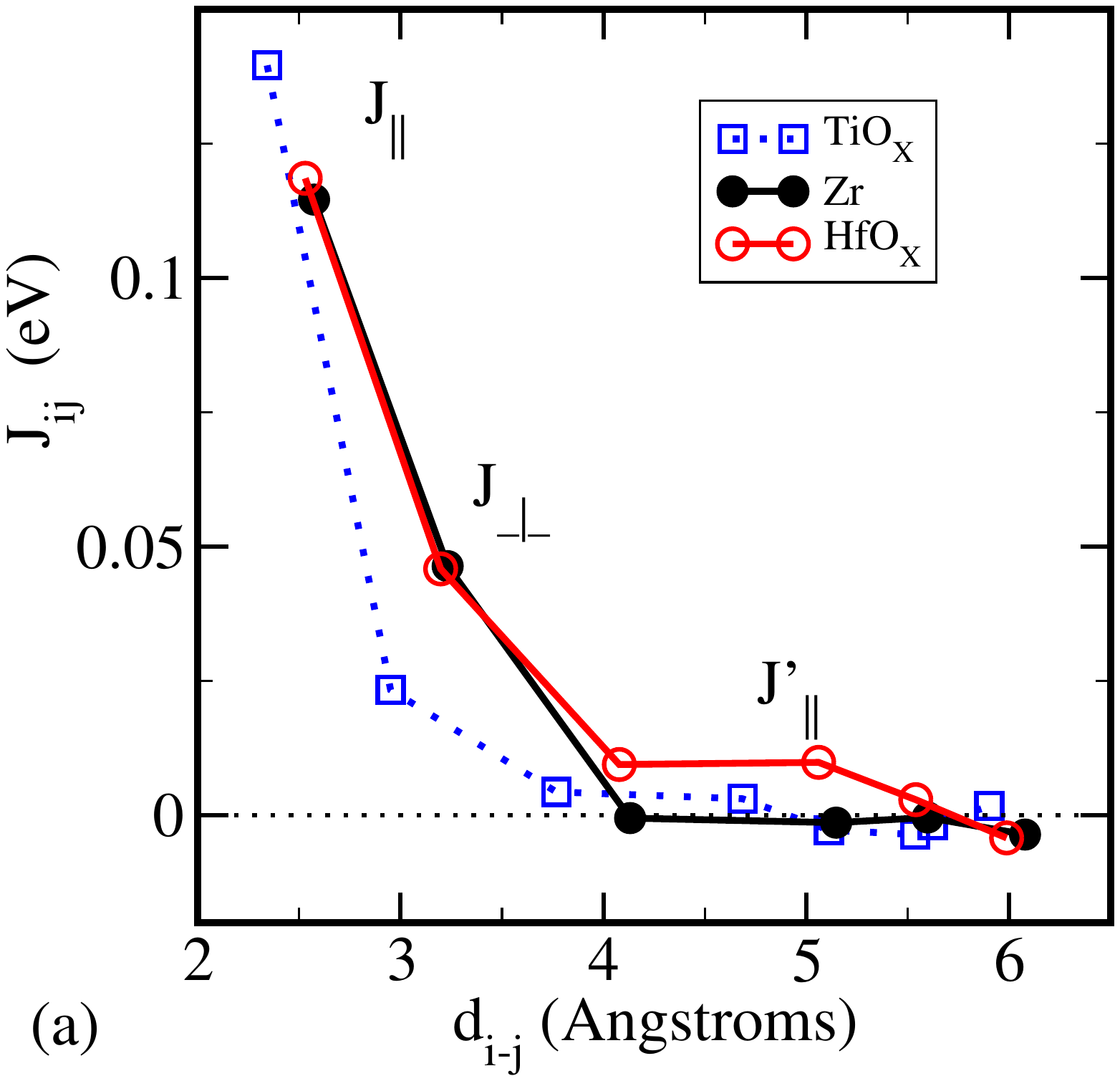}
\hspace{-0.5in}
\includegraphics[width=8.0cm,angle=0]{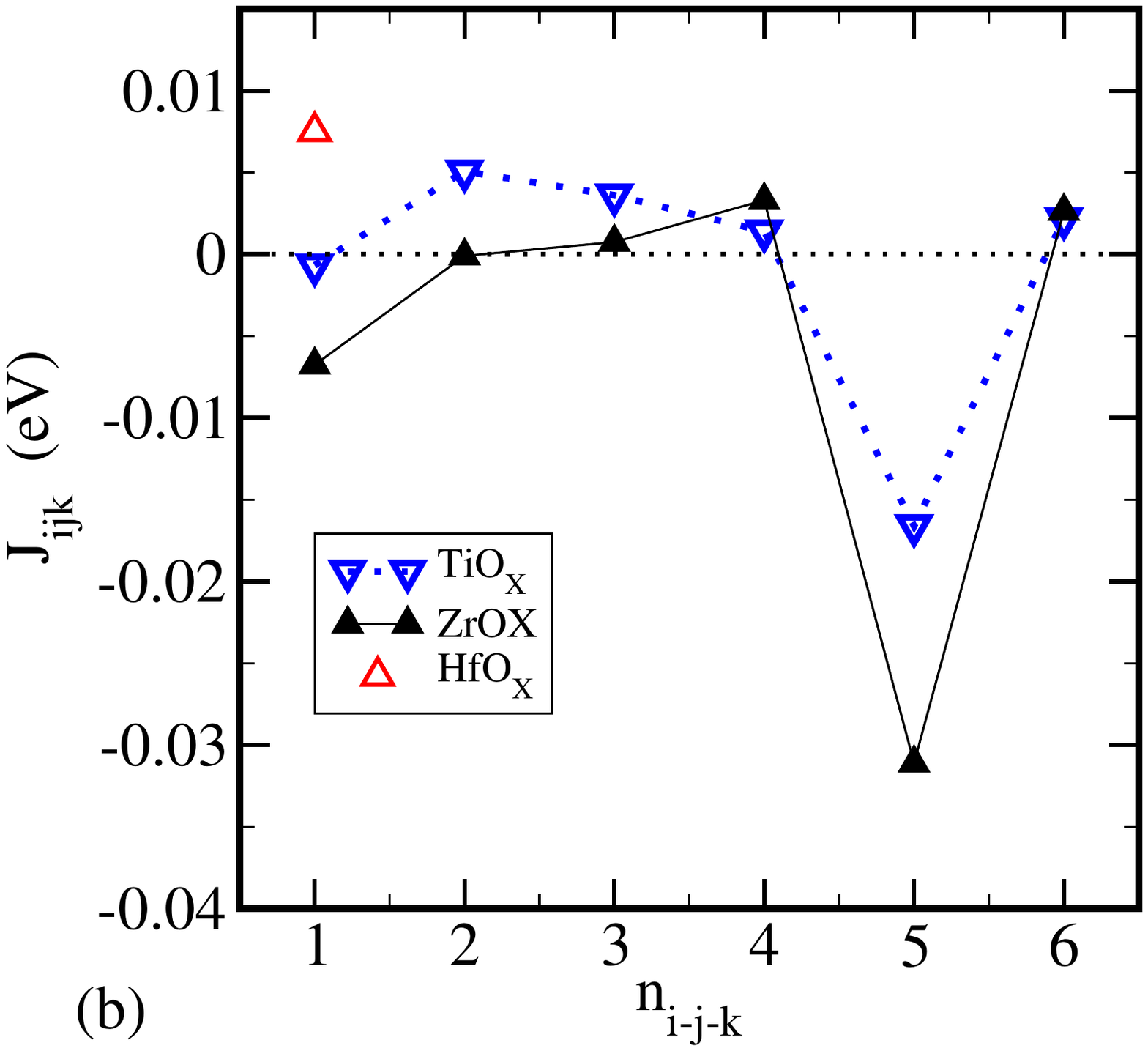}
\end{center}
\caption{Effective Cluster Interactions (ECI) for pair and
3-body interactions.  Solid black symbols indicate the
ZrO$_{X}$-fit which was used in the phase
diagram calculation reported here.  Open squares and down-pointing
triangles (blue online) indicate the results of a fit for the TiO$_{X}$
system. Open circles and open up-pointing triangles (red online) 
are from a fit for the HfO$_{X}$ system. 
(a) The first two pair-ECI are for
nearest-neighbor O-[~~] pairs that are
parallel- ($J_{\parallel}$) and perpendicular ($J_{\perp}$),
respectively, to c$_{Hex}$, and the 4'th nn pair-ECI
is the second-nn parallel to c$_{Hex}$
($J^{\prime}_{\parallel}$). Pairwise-ECI are
plotted as functions of inter-site separation.
(b) 3-body interactions are plotted as functions of the
index n$_{i-j-k}$ ~ which increases, nonlinearly,
as the area of triangle i-j-k increases. Positive pairwise
ECI imply a strong nn-pairwise O-[~~] attraction,
i.e. a strong nn-pairwise O-O repulsion.}
\label{fg:eci}
\end{figure}

Fitting was performed with the Alloy Theoretic Automated Toolkit (ATAT)
\cite{Axel2002a,Axel2002b,Axel2002c,Disclaimer} which automates most of the tasks
associated with the construction of a CE Hamiltonian.  A complete description
of the algorithms underlying the code can be found in
\cite{Axel2002b}. The zero- and point-cluster values were -0.421118 eV and
0.006221 eV, respectively.  The six pair and six 3-body ECI that
comprise the complete CE Hamiltonian are
plotted in Figs. \ref{fg:eci}a and \ref{fg:eci}b, respectively.
ECI for the isostructural TiO$_{X}$ (open symbols, blue online)
and HfO$_{X}$ (open symbols, red online) systems are also
plotted for comparison.
As expected, nearest neighbor (nn) O-O pairs
are highly energetic, and therefore strongly avoided; hence nn-pair
ECI are strongly $attractive$~ (ECI $\verb+>+0$, for O-[~~] nn pairs); but
beyond nn-pairs, the O-[~~] pairwise ECI are close to zero. The ratio
of magnitudes for nn-pair ECIs that are parallel- ($J_{\parallel}$) and perpendicular
($J_{\perp}$) to c$_{Hex}$, respectively, is
$J_{\parallel} / J_{\perp} \approx 2.5$. 
Note that the 4'th nn-pair ECI is the second-nn parallel to c$_{Hex}$, 
($J^{\prime}_{parallel}$) and  
$J^{\prime}_{parallel} / J_{\parallel} \approx 0.09$.

These results are similar to those presented in Ruban et al.
\cite{Ruban2010} although their effective pair interactions 
and ours are not identically defined.

Long-period superstructure (LPSS) phases were reported
\cite{Fehlmann1969,Yamaguchi1970} in samples with
with bulk compositions close to Zr$_{3}$O (essentially 
the $\alpha_{3} ^{\prime \prime}$ field in
Arai and Hirabayashi \cite{Arai1976}, their Fig. 9). 
Hence, it is reasonable to speculate that the CE-Hamiltonian 
might be like that in an axial next nearest neighbor Ising 
model (ANNNI-model), \cite{Bak1980} in which
one expecs $J_{\parallel}$~ and $J^{\prime}_{\parallel}$ to be opposite
in sign,and of comparable magnitudes
($0.3 \aplt -J^{\prime}_{\parallel}/J_{\parallel} \aplt 0.7$ \cite{Bak1980});
however, $J^{\prime}_{\parallel}/J_{\parallel} \approx 0.09$
(Figs. \ref{fg:eci}).

\section{Results}

\subsection{Ground-States}

The CE was used for a ground-state (GS) analysis that
included all configurations of [~~] and O in systems of
18 or fewer Zr-atoms (octahedral interstitial sites);
a total of $2^{18} = 262,144$~ structures (reduced by symmetry).
Five GS were identified in the range, $0 \leq X \leq 1/2$,
i.e. at X = 0, 1/6, 1/3, 5/12 and 1/2;
solid circles (black online) on the convex hull (solid line)
in Fig. \ref{fg:GS}.
The extension of the convex hull towards monoclinic
zirconia ($ZrO_2$) is also plotted in Fig. \ref{fg:GS}. The
CE-results suggest that all four VASP-GS in the
$\alpha Zr[~~]_{1-X}O_{X}$ subsystem are also GS of the Zr-O binary,
and that the maximum solubility of O in $\alpha$Zr$_{hcp}$~
is X$_{max} \approx 1/2$~ (higher than the
experimental value, X$~ \approx$~0.4).
Note that, the predicted CE-GS at Zr$_3$O$_2$~ is not
a GS for the VASP calculations (not a VASP-GS); hence the
VASP-predicted maximum solubility of O in Zr is X$_{max} \approx 0.5$.

The larger open squares (red online) in Figure \ref{fg:GS} are
CE-calculated values for the $\Delta E_{f}$~ that correspond
to the VASP calculations, and the smaller open squares (blue online)
are $\Delta E_{f}$~ for the remaining 262,144-96=262048 structures
in the GS analysis. The open diamond symbol (green online) indicates the
calculated formation energy for the P6$_3$22 structure for Zr$_3$O that
was originally proposed by Holmberg and Dagerhamn \cite{Holmberg1961};
this structure is also described in Table I. All space group determinations
were performed with the FINDSYM program. \cite{Disclaimer,FINDSYM}

~~~~\\
~~~~\\
~~~~\\
~~~~\\
~~~~\\
~~~~\\
~~~~\\

\pagebreak

\begin{longtable}[!t]{|c|c|c|c|l|}
\caption[]{Crystal structure parameters for
predicted ground-state phases in the $\alpha Zr[~~]_{1-X}O_{X}$~
system.  Cell constants are given in $\AA$.} \\ \hline
System       &     X      & Space Group       &Calculated cell            &Idealized              \\
             &  atomic    & IT number         & constants                 &Atomic                 \\
             & fraction O & Pearson Symbol    & ($\AA$)                   & Coordinates           \\ \hline \hline
Zr$_6$O      &   1/6      & R$\overline{3}$   &$a \approx \surd \overline{3} a_0$& O: 1/6, 1/6, 1/6  \\
             &            & 148               &$ = 5.5333$                & Zr: 3/4, 1/12, 5/12   \\
             &   1/7      & hP7               &$c \approx 3c_0 = 15.333$  & Zr: 11/12, 7/12, 1/4  \\
             &            &                   &                           & Zr: 1/12, 5/12, 3/4   \\
             &            &                   &                           & Zr: 1/4, 11/12, 7/12  \\
             &            &                   &                           & Zr: 5/12, 3/4, 1/12   \\
             &            &                   &                           & Zr: 7/12, 1/4, 11/12  \\ \hline \hline
Zr$_3$O      &   1/3      & R$\overline{3}$c  &$a \approx \surd \overline{3} a_0$& O: 1/6, 1/6, 1/6  \\
             &            & 167               &$ = 5.5671 $               & O: 2/3, 2/3, 2/3         \\
             &   1/4      & hP8               &$c \approx 3c_0 = 15.381$  & Zr: 3/4, 1/12, 5/12       \\
             &            &                   &                           & Zr: 11/12, 7/12, 1/4 \\
             &            &                   &                           & Zr: 1/12, 5/12, 3/4   \\
             &            &                   &                           & Zr: 1/4, 11/12, 7/12  \\
             &            &                   &                           & Zr: 5/12, 3/4, 1/12   \\
             &            &                   &                           & Zr: 7/12, 1/4, 11/12  \\ \hline \hline
Zr$_3$O      &   1/3      & P6$_3$22          &$a \approx \surd \overline{3} a_0$& O: 1/3, 2/3, 0    \\
             &            & 182               &$ = 5.5585 $               & O: 2/3, 1/3, 1/2      \\
             &   1/4      & hP8               &$c \approx c_0 = 5.1327$   & Zr: 1/3, 0, 0         \\
             &            &                   &                           & Zr: 0, 1/3, 0         \\
             &            &                   &                           & Zr: 2/3, 2/3, 0       \\
             &            &                   &                           & Zr: 2/3, 0, 1/2       \\
             &            &                   &                           & Zr: 0, 2/3, 1/2       \\
             &            &                   &                           & Zr: 1/3, 1/3, 1/2     \\ \hline \hline
Zr$_{12}$O$_5$ &  5/12      & R$\overline{3}$   &$a \approx \surd \overline{3} a_0$& O: 1/12, 1/12, 1/12  \\
             &            & 148               & $ = 5.5568 $              & O: 1/4, 1/4, 1/4      \\
             &  5/17      & hP17              & $c \approx 3c_0 = 30.861$ & O: 1/2, 1/2, 1/2         \\
             &            &                   &                           & O: 2/3, 2/3, 2/3         \\
             &            &                   &                           & O: 11/12, 11/12, 11/12   \\
             &            &                   &                           & Zr: 1/8, 11/24, 19/24    \\
             &            &                   &                           & Zr: 1/24, 17/24, 3/8     \\
             &            &                   &                           & Zr: 23/24, 7/24, 5/8     \\
             &            &                   &                           & Zr: 21/24, 13/24, 5/24   \\
             &            &                   &                           & Zr: 19/24, 1/8, 11/24    \\
             &            &                   &                           & Zr: 17/24, 3/8,  1/24    \\
             &            &                   &                           & Zr: 5/8, 23/24,  7/24    \\
             &            &                   &                           & Zr: 13/24, 5/24, 21/24   \\
             &            &                   &                           & Zr: 11/24, 19/24, 1/8    \\
             &            &                   &                           & Zr: 3/8, 1/24, 17/24     \\
             &            &                   &                           & Zr: 7/24, 5/8, 23/24     \\
             &            &                   &                           & Zr: 5/24, 7/8, 13/24     \\ \hline \hline
Zr$_2$O      &   1/2      & P$\overline{3}$1m &$a \approx \surd \overline{3} a_0$&0, 0, 0         \\
             &            & 162               & $ = 5.5501 $              & O: 1/3, 2/3, 1/2      \\
             &   1/3      & hP9               &$c \approx c_0 = 5.1572$   & O: 2/3, 1/3, 1/2      \\
             &            &                   &                           & Zr: 0, 1/3, 3/4       \\
             &            &                   &                           & Zr: 1/3, 1/3, 1/4     \\
             &            &                   &                           & Zr: 1/3, 0, 3/4       \\
             &            &                   &                           & Zr: 2/3, 0, 1/4       \\
             &            &                   &                           & Zr: 2/3, 2/3, 3/4     \\
             &            &                   &                           & Zr: 0, 2/3, 1/4       \\ \hline \hline

\end{longtable}

\pagebreak

\begin{figure}[!htbp]
\begin{center}
\includegraphics[width=20.cm,angle=0]{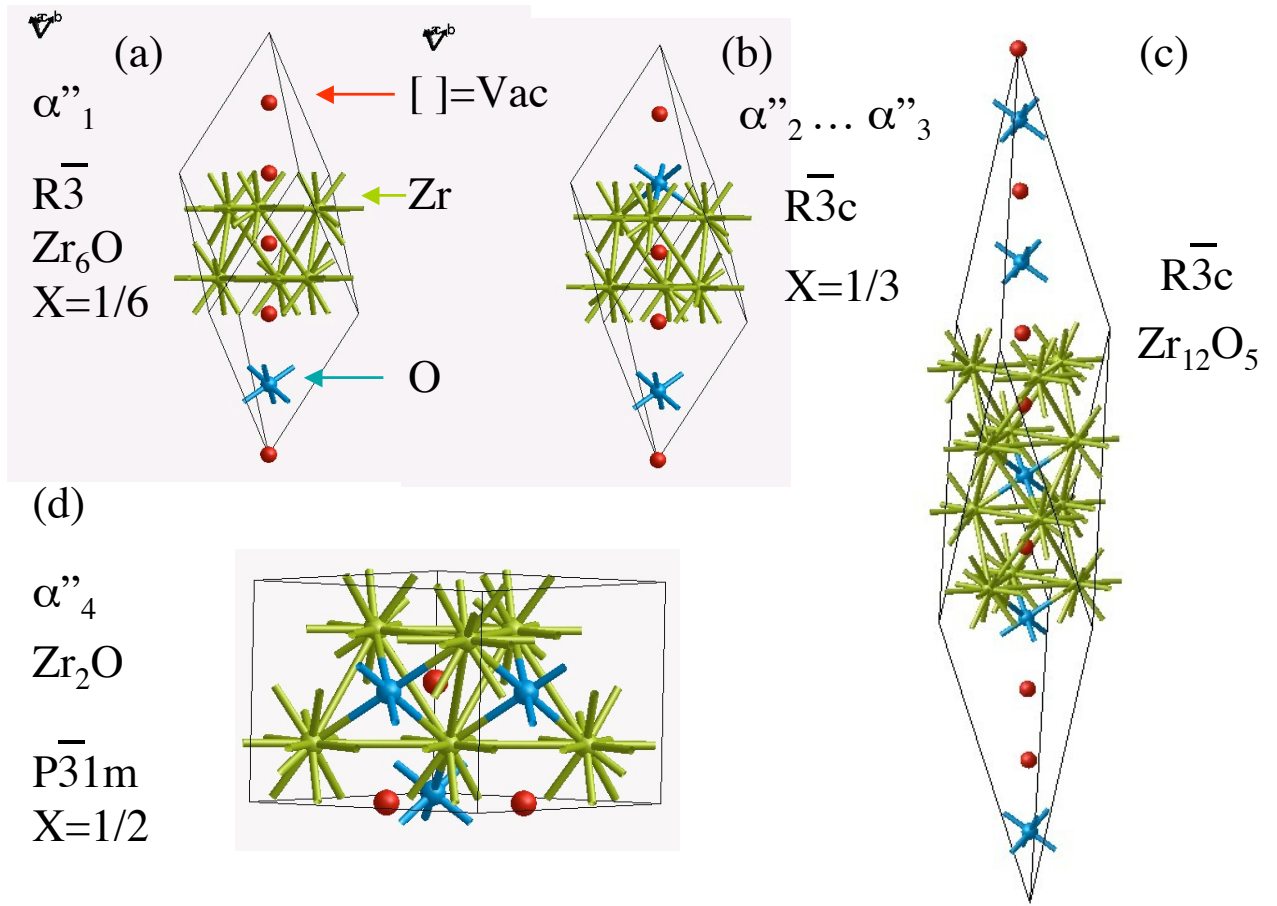}
\end{center}
\vspace{-0.5in}
\caption{Idealized crystal structures of the four
cluster-expansion-predicted suboxide ground-states:
(a) Zr$_6$O; (b) Zr$_3$O; (c) Zr$_{12}$O$_5$; (d) Zr$_2$O.
Spheres connected by bond-sticks (yellowish-green
online) represent Zr. Isolated spheres with bond-sticks
(blue online) represent oxygen. Isolated spheres
(red online) represent vacant octahedral sites.
}
\label{fg:XalGS}
\end{figure}
\pagebreak

Crystal structures of the VASP- and CE-GS in Zr-ZrO are
described in Table I and their idealized structures
are drawn in Figures \ref{fg:XalGS} a-d:
Zr is represented by spheres connected with bond-sticks
(yellowish-green online); O is represented by isolated
spheres with bond-sticks (blue online); and [~~] are
represented by isolated spheres (red online).

Various low-T ordered structures have been reported,
\cite{Yamaguchi1968,Yamaguchi1970,Hirabayashi1972,Hirabayashi1974,Hashimoto1974,Arai1976,Sugizaki1985}
with the most recent review by Sugizaki et al. \cite{Sugizaki1985};
who used neutron diffraction
to study short-range order in ZrO$_{0.3}$~ solid solutions.
Their Figs. 1a-c presented representations of three
ordering schemes that were observed within different
homogeneity ranges:
(a) ZrO$_x$ at X $\aplt 1/3$ (P321);
(b) ZrO$_y$ at 1/3~ $\aplt ~X~ \aplt 0.4$ (P6$_3$22);
(c) ZrO$_z$ near the solubility limit X~$\approx ~ 0.4$ (P$\overline{3}1m$).
Space groups for these idealized structures were not reported by
Sugizaki et al. \cite{Sugizaki1985}; they were determined in this work with
the FINDSYM program. \cite{FINDSYM}
Comparing structures (a)-(c) above to the results of this work:
(a) VASP calculations indicate that this structure is clearly not a GS;
(b) is the P6$_3$22 structure shown as an open diamond (green online)
in Fig. \ref{fg:GS}, its formation energy is very close, but higher
than, the VASP-GS at X=1/3;
(c) is the same P$\overline{3}1m$ structure as the VASP-GS at
X=1/2.

\subsubsection{Zr$_6$O, X=1/6, $\alpha_1^{\prime \prime}$}

The structure of Zr$_6$O is thought
to be isomorphic to that of Hf$_6$O and Ti$_6$O
\cite{Arai1976,Abriata1986}:
a$\approx \surd \overline{3}a_0$;
c$\approx$c$_0$; Z=3 (a$_0$ and c$_0$ are the cell constants
of the disordered P6$_3$mmc alloy).\cite{Hirabayashi1972}
This is also the VASP-GS at X=1/6, Fig. \ref{fg:XalGS}(a) and  Table 1.

\subsubsection{Zr$_3$O, X=1/3, $\alpha_2 ^{\prime \prime}....\alpha_3 ^{\prime \prime}$}
Based on X-ray diffraction studies,
Holmberg and Dagerhamn \cite{Holmberg1961} proposed a
P6$_3$22 structure (open diamond, green online, in
Fig. \ref{fg:GS}) with a$\approx \surd \overline{3}a_0$
and c$\approx c_0$ for a sample with X$\geq 0.26$.
Based on single crystal neutron diffraction studies
Yamaguchi \cite{Yamaguchi1968} reported X-ray, electron and neutron
diffraction data on samples in the range ZrO$_{0.18}$-ZrO$_{0.30}$
(1/5$ \leq X \leq 3/7$) and listed atomic coordinates for a
"P3c1" structure with $a \approx \surd \overline{3}a_0$, c$\approx 3c_0$.
Yamaguchi \cite{Yamaguchi1968} also reported confirmation of the P6$_3$22
structure in the composition range 0.33 $\verb+<+ ~X~ \verb+<+ 0.4~$
(1/2 $\verb+<+ X \verb+<+ 2/3$).
The FP results presented here suggest that the
VASP-GS at X=1/3 is the R$\overline{3}$c structure depicted in
Figure \ref{fg:XalGS} (b). The calculated energy-difference between
these two structures is only 0.006 eV, and this difference is probably
within DFT error, but the precision of these calculations is
sufficient to recognize the R$\overline{3}$c structure as the
VASP-GS.

\subsubsection{Zr$_{12}O_5$, X=5/17}

This structure does not correspond to any reported phase,
and because it is predicted to disproportionate at T$ \geq 20$K.
It is not expected to be observed experimentally.

\subsubsection{Zr$_2$O, X=1/2, $\alpha_4 ^{\prime \prime}$}

The only Zr$_2$O~ structure listed in Pearson \cite{Pearson} is
cubic, and the apparent solubility limit of X$ \approx$ 0.4, rather
than X=1/2, which suggests that the VASP calculations may underestimate
the stability of monoclinic ZrO$_2$, and therefore finds the
GS tieline between the P$\overline{3}$1m GS at X=1/2 and
monoclinic ZrO$_2$, rather than between the R$\overline{3}$c GS
at X=1/3 and monoclinic ZrO$_2$.
Another possibility is that the experimentally measured
low-temperature equilibrium between Zr-suboxides and monoclinic ZrO$_2$
was measured at too low a fugacity of oxygen to stabilize the
P$\overline{3}$1m phase at X=1/2.
As one expects from the ECI (Fig. \ref{fg:eci}),
there are no O-O nn pairs
in the VASP-GS P$\overline{3}$1m structure, or in any of the four
structures with formation energies within 0.01 eV
(right panel Fig. \ref{fg:GS}).

\subsection{The Phase Diagram}

\begin{figure}[!htbp]
\begin{center}
\hspace{-0.2in}
\includegraphics[width=7.0cm,angle=0]{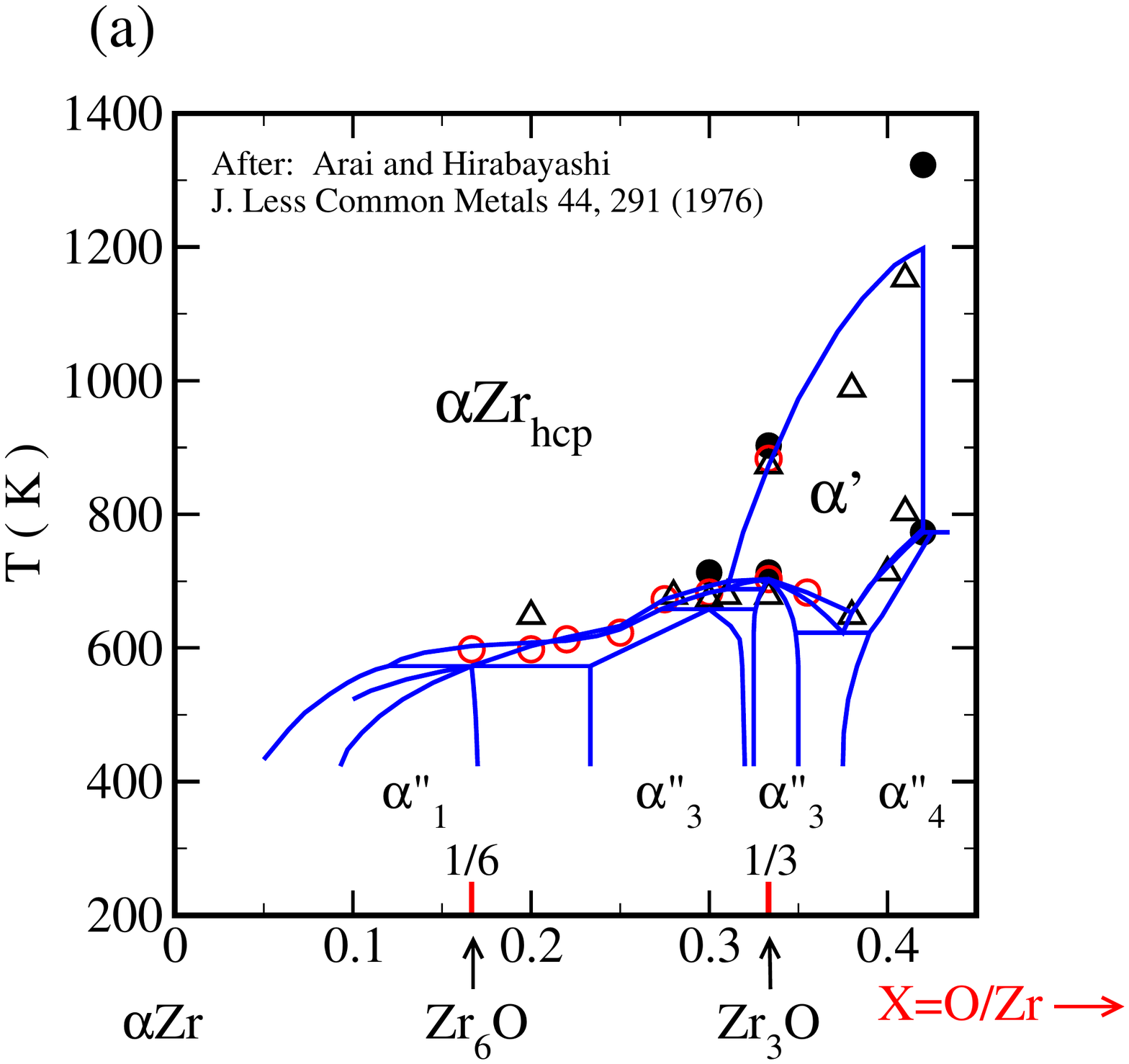}
\hspace{-0.2in}
\includegraphics[width=9.0cm,angle=0]{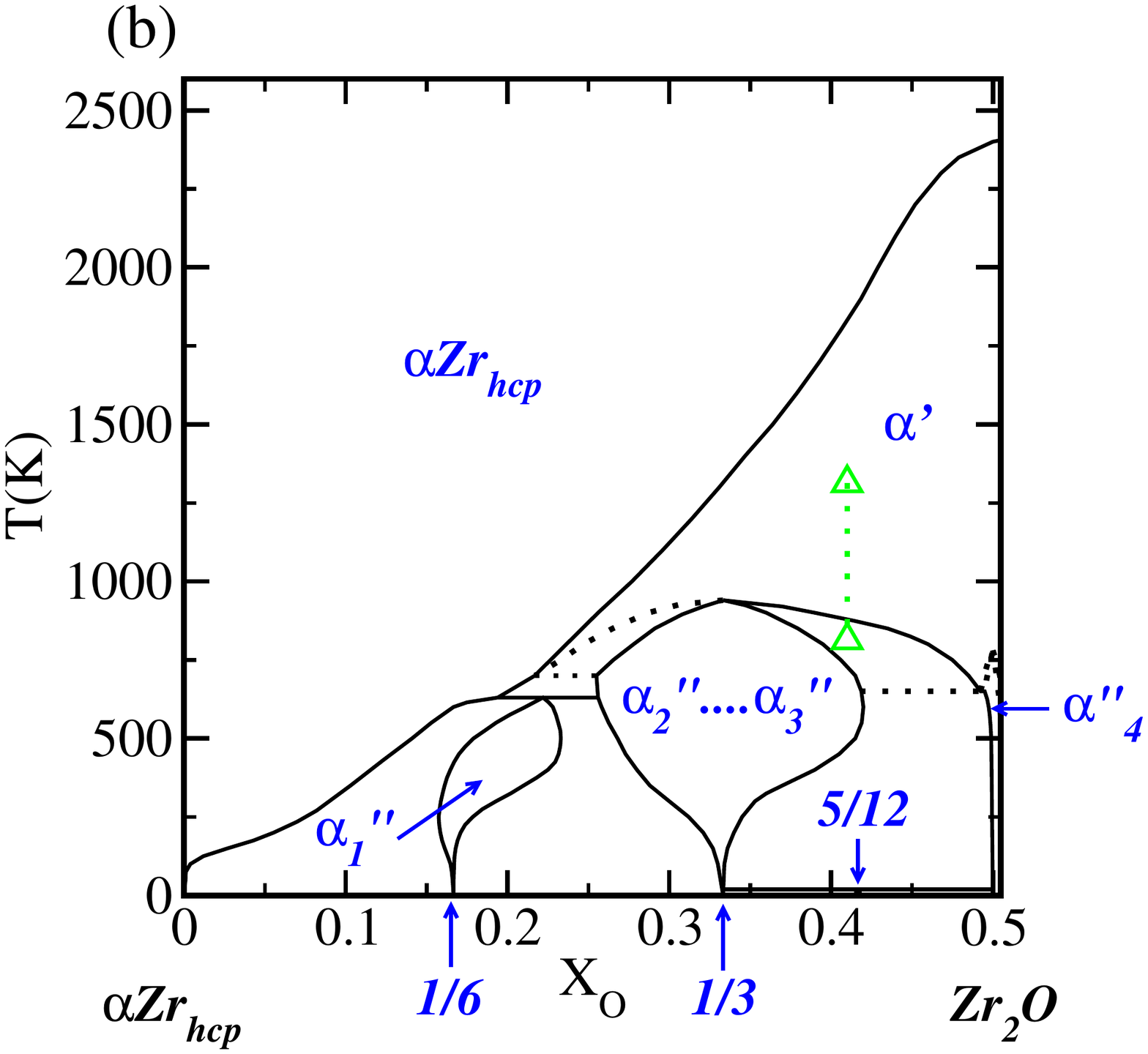}
\end{center}
\caption{Comparison of experimental and calculated
phase diagrams for the system $\alpha Zr[~~]_{1-X}O_{X}$:
(a) a combination of the  "transformational diagram"
(symbols) and the "tentative diagram" (solid lines) in Arai and
Hirabayashi \cite{Arai1976} (their Figs. 1 and 9, respectively);
(b) the diagram calculated from this work
(dotted phase boundaries are less precisely determined than
solid boundaries).
Note the different results for
$0.25 \aplt X \aplt 0.42$~ and $420K \aplt T \aplt 725K$.
}
\label{fg:XT}
\end{figure}

A first principles phase diagram (FPPD) calculation was performed
with grand canonical Monte Carlo (MC) simulations using the
emc2 code which is part of the ATAT
package \cite{Axel2002a,Axel2002b,Axel2002c}. Input parameters for
emc2 were: a simulation box with at least 1568 octahedral sites 
(15x15x6 supercell);
2000 equilibration passes; 2000 Monte Carlo passes.
The predicted phase diagram is shown in Figure \ref{fg:XT}.
Most phase boundaries were determined by following order-parameters of
the various ordered phases as functions of X and T; here
order parameters are defined such that they are unity
in a specified GS-phase, zero in the disordered phase, and 
typically some non-zero value in ordered phases other
than their specified GS.  
Dotted boundaries are used to acknowledge uncertainties
in phase boundary determinations.

\pagebreak

\subsection{The Intermeadiate Temperature $\alpha ^{\prime}$-Phase }

\begin{figure}[!htbp]
\begin{center}
\includegraphics[width=15.cm,angle=0]{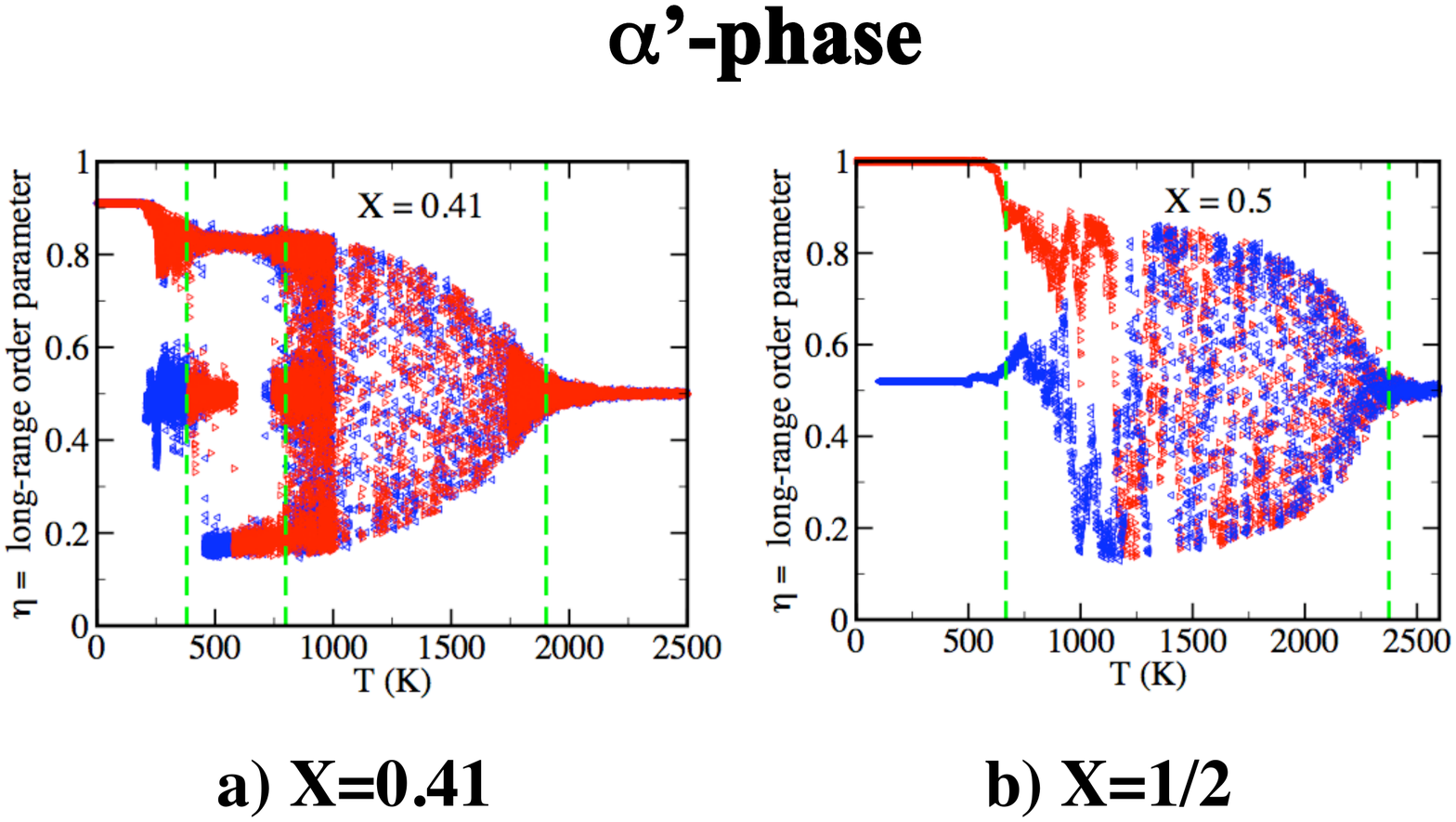}
\end{center}
\vspace{-1.0in}
\caption{Calculated order-parameter vs temperature
curves for: a) X=0.41; b) X=1/2. Heating simulations are 
indicated by right-pointing triangles (red online) and 
cooling simulations are represented by left-pointing triangles
(blue online).  As observed experimentally, there is a 
two-step disordering process on heating.
}
\label{fg:AprimeX}
\end{figure}

As observed experimentally in samples with X=0.41, \cite{Hirabayashi1974}
(up-pointing triangles, green online, Fig. \ref{fg:XT})  a two-step
order-disorder process is predicted for $0.25 \aplt X \aplt 0.5$ 
Figures \ref{fg:AprimeX}.
The data reported in Hirabayashi et al. \cite{Hirabayashi1974} appear to
indicate that both order-disorder transitions are second-order (continuous)
in character, at least at X=0.41, but the calculations reported here
suggest that the lower-T transition is strongly first-order (at least
at X=1/2) while the higher-T transition is continuous.

\pagebreak

The simulated intermediate-temperature $\alpha ^{\prime} $-phase 
crystal structure was determined by symmetry analysis, using
the ISODISTORT program \cite{Disclaimer,ISODISTORT}. There are
two plausible transition paths from the P6$_3$/mmc high-T disordered phase
to the P$\overline{3}$1m GS: \\
\begin{itemize}
\item{(1) P6$_3$/mmc $\rightarrow$ P6$_3$/mcm $\rightarrow$ P$\overline{3}$1m,
$K_{1}$ irrecucible representation, (-1,-2,0),(2,1,0),(0,0,1) basis;}
\item{(2) P6$_3$/mmc $\rightarrow$ P$\overline{3}$m1 $\rightarrow$ P$\overline{3}$1m,
$\Large \Gamma_{3}^+$ irreducible representation, (0,-1,0),(1,1,0),(0,0,1) basis.}
\end{itemize}
Path (1) can be ruled out because it requires a first-order 
P6$_3$/mmc $\rightarrow$ P6$_3$/mcm transition, with unit-cell expansion 
along both $a_{Hex}$ axes, which neither experiment nor computation supports. 

\begin{table}  [!t]
\caption{Atomic positions in P$\overline{3}$m1 (IT 164) $\alpha ^{\prime}$ crystal \\
structue: $a \approx a_0 \approx 3.32 \AA$; $c \approx c_0 \approx 5.14000 \AA$;
$^*$ X=O/Zr.} 
\begin{tabular}{|c|c|c|c|c|c|} \hline  \hline
 Atom  & Wyckoff site &   x   &   y   &        z       & occupancy             \\ \hline \hline
 Zr    &      2d      & 1/3   &  2/3  & $\approx 1/4$  &    1                  \\ \hline
 O$_1$ &      1a      &  0    &   0   &        0       & $x_{O1} \verb+<+ 1/2$ \\
 O$_2$ &      1b      &  0    &   0   &       1/2      & $2X^* - x_{O1}$       \\  \hline \hline 
\end{tabular} 
\label{tb:P-3m1}
\end{table}

\begin{figure}[!htbp]
\begin{center}
\includegraphics[width=10.0cm,angle=0]{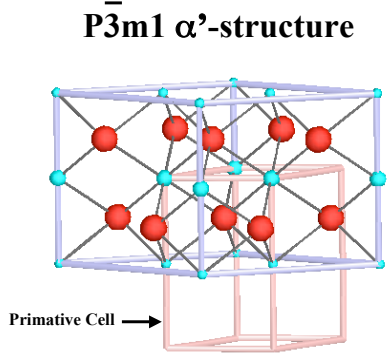}
\end{center}
\vspace{-0.5in}
\caption{Average P$\overline{3}$m1 structure of the 
$\alpha ^{\prime}$-phase. Small and intermediate sized spheres 
(blue online) represent less- and more oxygen-rich 
oxygen:vacancy-sites (O:[~~]-sites),
respectively.  Larger spheres (red online) represent Zr atoms. 
More- and less O-rich O:[~~]-sites segregate into alternating 
layers perpendicular to $c_{Hex}$; reflecting nearest neighbor
O-O avoidance. 
}
\label{fg:AlphaPrime}
\end{figure}

\begin{figure}[!htbp]
\begin{center}
\includegraphics[width=15.0cm,angle=0]{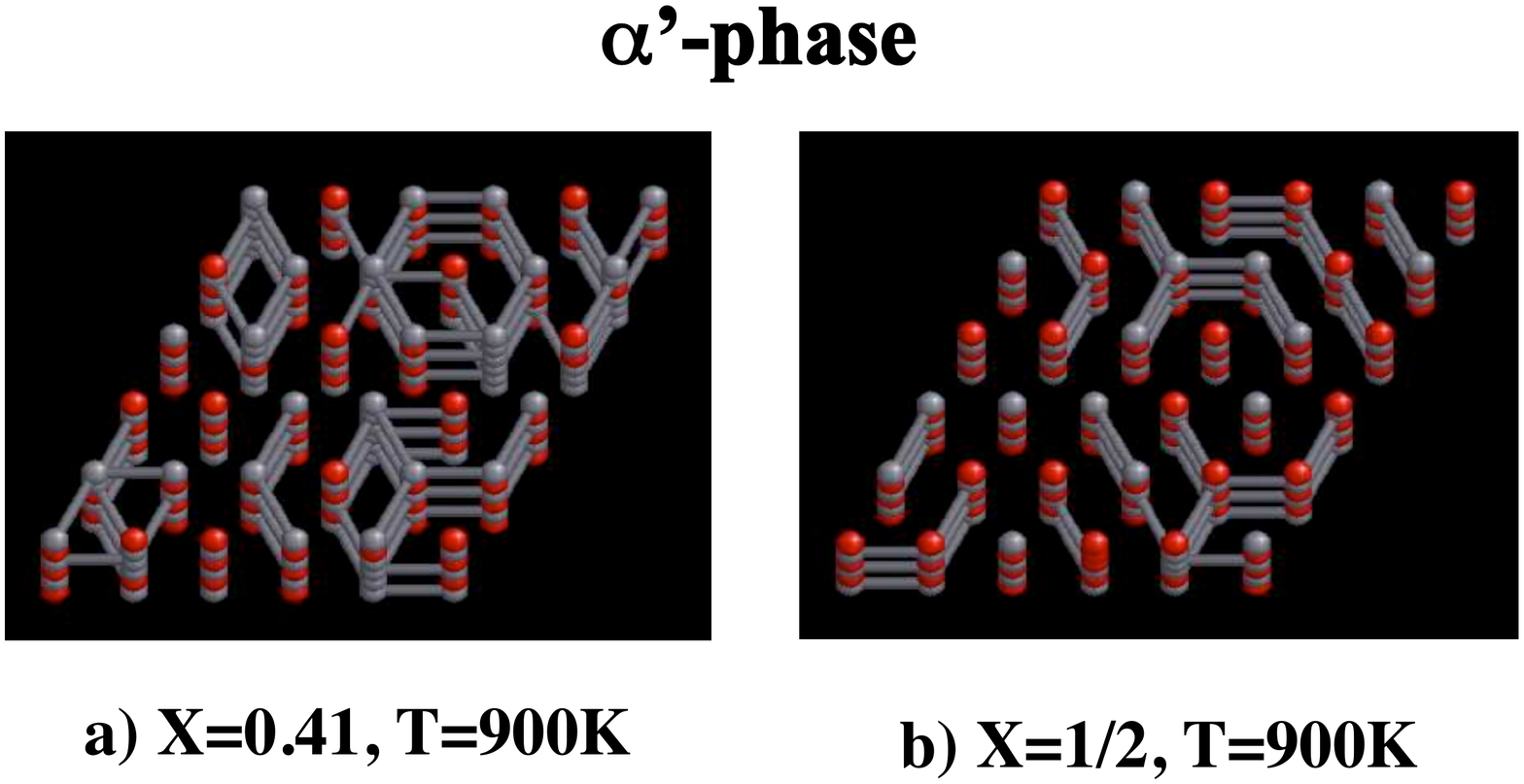}
\end{center}
\vspace{-1.0in}
\caption{Simulated O:[~~]-sites (red:gray online, respectively)
distributions at: (a) X=0.41, T=900K; and (b) X=1/2, T=900K.
For clarity, Zr-atoms are omitted and a reduced (6x6x3) supercell
were used.
At X=0.41 no O-O nn pairs are evident parallel to $c_{Hex}$.
At X=1/2, almost all nn pairs parallel to $c_{Hex}$~ are O-[~~], although 
two columns (first row, columns 4 and 5) have some O-O nn pairs), 
while perpendicular to $c_{Hex}$~ there are many more 
O-O and [~~]-[~~] nn pairs.  
}
\label{fg:AprimeDist}
\end{figure}

Path (2) permits a continuous P6$_3$/mmc $\rightarrow$ P$\overline{3}$m1 
transition, as observed experimentally and supported computationally. 
The $average$~ $\alpha ^{\prime}$, P$\overline{3}$m1 structure is 
described in Table \ref{tb:P-3m1} and depicted in
Fig. \ref{fg:AlphaPrime}; where $partially~ occupied$ O:[~~]-sites are represented
by relatively smaller and larger spheres (blue online). The precise
occupations of sites O$_{1}$~ and O$_{2}$~ can be written as
$\chi$~ and  $2X - \chi$, respectively; 
where $\chi \verb+ < + 1/2$~ is the O-occupancy of site O$_{1}$,
and  X=O/Zr; i.e. at X=0.41 and $\chi = 0.22$~ then $2X- \chi = 0.60$.
With respect to space-group determination, the only requirement is
that the occupancy of O$_{1}$~ must be different from that
of O$_{2}$. 
The P$\overline{3}$m1 structure is clearly consistent with the computational 
results shown in Figures \ref{fg:AprimeDist}a and \ref{fg:AprimeDist}b. 
The O:[~~]-distributions (online O=red, [~~]=gray) in these figures
were simulated on reduced (6x6x3) supercells by cooling from 1000K to 900K.
For clarity Zr-atoms are omitted to highlight
the strong preference for O:[~~]-ordering along $c_{Hex}$; i.e.
strong O-O nn avoidance along $c_{Hex}$.  In the $average$~ P$\overline{3}$m1 structure
this leads to alternating nn-layers, $\perp ~ c_{Hex}$~ that are 
relatively O-rich and O-poor ([~~]-rich). Visually, this statistical 
difference is obscured in the simulation snapshots (Figures \ref{fg:AprimeDist}a 
and \ref{fg:AprimeDist}b) because one has: discrete O and [~~]; O:[~~]-disorder; 
and antiphase boundaries.

\section{Discussion}

\subsection{Comparison of Calculated and Experimental Phase Diagrams}

The main differences between the FPPD presented here and the
"tentative phase diagram" in Arai and Hirabayashi \cite{Arai1976}
(Fig. \ref{fg:XT}a; their Fig. 9) are with respect to their
representations of broad homogeneity ranges for three ordered phases
in the range $0.25 \aplt X \aplt 0.42$ and $420K \aplt T \aplt 725K$.
In this range, Arai and Hirabayashi report three low-T
ordered phases, $\alpha_{2} ^{\prime \prime}$,
$\alpha_{3} ^{\prime \prime}$, and $\alpha_{4} ^{\prime \prime}$; whereas the FPPD has only two;
note that the predicted GS at X=5/12 disproportionates at
T$ \approx 20K$.  Also, the FPPD-predicted
$\alpha ^{\prime}$-phase field is significantly larger than the corresponding
field in Fig. \ref{fg:XT}a, and at X=0.41 the
$\alpha ^{\prime}$-$\alpha Zr_{hcp}$ transition is predicted to occur
$\approx 500K$ higher than experiment suggests, Fig. \ref{fg:AprimeX}a.
Typically, FPPD calculations overestimate order-disorder transition
temperatures especially when, as here, the excess vibrational
contribution to the free energy \cite{VIB} is ignored; so it is not surprising
that agreement between experiment and theory is not close for the
$\alpha ^{\prime} \leftrightharpoons \alpha Zr_{hcp}$~ order-disorder transition.
Note however, that the maximum temperatures for stabilities of
phases other than $\alpha ^{\prime}$~ are roughly equal to those shown
in Fig. \ref{fg:XT}a.

\subsection{Long-Period Superstructures at X$\approx$1/3}

Based on X-ray, neutron, and electron scattering data,
Fehlmann et al. \cite{Fehlmann1969} and
Yamaguchi and Hirabayashi \cite{Yamaguchi1970} reported
a variety of long-period superstructures (LPSS) in samples
with bulk compositions X$\approx$1/3 (the $\alpha_{3} ^{\prime \prime}$
field, Fig. \ref{fg:XT}a) that were
subjected to various heat treatments.
The FPPD calculation presented here does not predict 
LPSS fields, but a similar calculation for HfO$_{X}$~ appears
to predict Devil's Staircases of ordered phases at
Hf$_{3}$O and Hf$_{2}$O. \cite{Burton2011}
In an ANNI-model like Hamiltonian, one expects,
$0.3 \aplt -J_{\parallel}'/J_{\parallel} \aplt 0.7$,
however, the 12-pair fit which includes $J_{\parallel}'$
yields $J_{\parallel}$~ and $J_{\parallel}'$ with the same sign
and $J_{\parallel}' \approx J_{\parallel}/10$.
Physically, the fitted values for $J_{\parallel}$~ and $J_{\perp}$~ are
easy to rationalize in terms of O-O nn-repulsion, and this argues
against stable LPSS phases, unless they are stabilized by competition
between higher-order interactions; e.g. 3'rd and further nn-pair-ECI 
or multiplet interactions.  In fact, FPPD calculations for the HfO$_{X}$~
system, which has a CE Hamiltonian very similar to that for ZrO$_{X}$, 
indicate a Devis's Staircase of LPSS phases at Hf$_{3}O$. \cite{Burton2011}

\section{Conclusions}

Ground-State ordered phases are predicted at
X=0, 1/6, 1/3, 5/12 and 1/2, but the one at X=5/12 is
predicted to disproportionate at T$\approx 20K$, hence it
is not expected to be observed experimentally.
In the range $0.25 \aplt X \aplt 0.5$, in which Arai and Hirabayashi \cite{Arai1976}
report three phases 
($\alpha_{2} ^{\prime \prime}, ~ \alpha_{3} ^{\prime \prime}$ and $\alpha_{4} ^{\prime \prime}$)
only two are predicted; i.e. the phase fields that Arai and Hirabayashi \cite{Arai1976}
draw for $\alpha_{2} ^{\prime \prime}$ ~ and $\alpha_{3} ^{\prime \prime}$~ are predicted to be a
single-phase solid solution.
Figure 1a clearly indicates that a zeroth order model
for octahedral interstitial O:[~~]-ordering is one in which first- and
second-nn pairwise interactions ($J_{\parallel}$ and
$J_{\perp}$, respectively) strongly favor O-[~~] nn-pairs;
i.e. O-O nn-pairs are highly unfavorable, and
$J_{\parallel} / J_{\perp} \approx 2.5$.
Including $J^{\prime}_{\parallel}$~ in the ECI fit does not yield an
ANNNI-like \cite{Bak1980} CE-Hamiltonian;
however, recent FPPD calculations for the HfO$_{X}$~ 
system, \cite{Burton2011} (the HfO$_{X}$-CE is 
very similar to the ZrO$_{X}$-CE) predict Devis's Staircases 
of LPSS phases at Hf$_{3}$O and Hf$_{2}$O.

The most probable transition path (on cooling) for O-rich solutions, 
$X \apgt 0.4$ is P6$_3$/mmc $\rightarrow$ P$\overline{3}$m1 $\rightarrow$ P$\overline{3}$1m;
hence the average $\alpha ^{\prime}$-structure has P$\overline{3}$m1 symmetry with
alternating O-rich and [~~]-rich layers $\perp ~ c_{Hex}$.


\subsection{Submitted as a "Full Paper" J. Phys. Soc. Japan Wed, 27 Apr 2011}


\begin{thebibliography}{999}
\bibitem{Disclaimer}
Disclaimer: the use of a specific software package should
not be misinterpreted as implying a NIST endorsement of
that package.
\end{thebibliography}
\end{document}